\documentstyle[11pt]{article}
\begin{document}
\centerline{\Large\bf Relativistic quantum protocols:}
\vskip 2mm
\centerline{\Large\bf ``Bit Commitment'' and ``Coin Tossing''}
\vskip 3mm
\centerline{S.N.Molotkov and S.S.Nazin}
\centerline{\sl\small Institute of Solid State Physics 
of Russian Academy of Sciences}
\centerline{\sl\small Chernogolovka, Moscow District, 142432, Russia}
\vskip 3mm

\begin{abstract}
The relativistic quantum protocols realizing the bit commitment
and distant coin tossing schemes are proposed. The protocols are based
on the fact that the non-stationary orthogonal extended quantum states
cannot be reliably distinguished if they are not fully accessible
for the measurement. As the states propagate from the domain controlled 
by one of the user to the domain accessible for the measurements performed
by the other user, they become reliably distinguishable for the second user.
Important for the protocol are both the quantum nature of the states and the
existence of a finite maximum speed of the signal propagation imposed by 
the special relativity.
\end{abstract} 
\vskip 3mm

\section{Introduction}
Many cryptographic problems reduce to a number of primitive cryptographic 
exchange protocols, such as the secret key distribution [1--3], bit 
commitment [4--6], and distant coin tossing [7] protocols. The bit commitment
protocol is stronger than the distant coin tossing one in the sense that a
distant coin tossing protocol can be formulated on the basis of a bit 
commitment protocol.

Informally, the bit commitment protocol is usually formulated in the following 
way. The protocol involves two participants (users) called A and B. At the
commitment stage the user A chooses the value of a secret bit (0 or 1)
and sends some information about his choice to user B in such a way 
that using the information provide by user A the user B cannot reliably 
determine the secret bit value chosen by A. To be more precise, in the ideal 
case the probability for user B to correctly identify the bit value chosen by 
user A is exactly 1/2 (i.e. it is equal to the probability of simply guessing 
the bit value) regardless of whether or not he uses the information supplied
by user A. Then at the disclosure stage the user B can ask user A to provide him
the rest information on the value of the chosen secret bit so that in the ideal 
case the user B reliably recovers the secret bit value. In addition, there 
should be no possibility for user A to change his mind and alter the chosen 
bit value after the commitment stage and before the disclosure state without
being caught by user B.

The distant coin tossing protocol is formulated in the following way.
The two distant users A and B, who do not trust each other and can employ
any physically realizable opportunities to cheat, should exchange appropriate 
information so that at the end of the protocol (in the ideal case with the
unit probability) they accept the arising bit as an honest lot. 
If the users have only access to the classical communication channel, 
the problem can even seem unsolvable. 

Obviously, realization of any bit commitment protocol can be used to construct
a distant coin tossing protocol. To achieve this purpose, user B can try and 
guess the bit chosen user A after the commitment stage but before the 
disclosure stage (remember that after the disclosure stage the secret bit 
chosen by user A is publicly known). The user B wins if he guesses the bit 
value chosen by user A and loses otherwise.

The following protocol is sometimes described as a simple example of the bit
commitment protocol. User A writes down the chosen bit value on a paper sheet 
and places it into a safe which is then sent to user B (commitment stage)
without the key which is only given to user B at the disclosure stage.
In spite of the simplicity of the above example, it contains all the basic 
features of the protocols based on the classical information carriers.
In this example, the user B obtains the complete rather than partial 
information on the secret bit at the commitment stage. Therefore, the laws 
of nature do not prohibit the user B to learn the secret bit value even before
the disclosure stage if he has access to sufficient technical resources.
A similar situation takes place in the protocols based on the computational
complexity of some trap-functions (e.g., discrete logarithm) [8]. In the 
protocols of that type user A announces to user B the value $y$ (where 
$y=a^b\mbox{mod}\mbox{ }p$; $a,p$ are known in advance, and the parity of $b$ 
is the secret bit value). In principle, the supplied information on bit $b$ 
(i.e., the value of function $y$) is sufficient to learn the secret bit by
calculating the discrete logarithm. However, all available classical numerical 
algorithms require exponentially large computational resources (although  
it was never proved that there exists no more efficient 
polynomial classical algorithm for this problem).

For the case where A and B can only exchange information through the classical 
communication channel, the problem was solved in Ref. [8]. Strictly speaking,
the protocol proposed in Ref. [8] is not secure against cheating by one of the 
users since it is based on the unproved computational complexity of the 
discrete logarithm problem [8].
 
In these protocols the user B is given the complete rather than partial
information on the secret bit already at the commitment stage. Therefore,
in principle, user B can learn the secret bit even before the disclosure 
stage, for example, by using the quantum computer [9,10] (which is currently,
however, very far from the experimental realization).

Employing only the classical (non-relativistic) objects as the information 
carriers, it is impossible to construct an unconditionally secure bit 
commitment protocol (whose security is based on the fundamental laws of 
nature only rather than current technical limitations) where only ``part'' of 
a classical object (for example, a spatially extended signal which is only 
partly accessible to user B before the disclosure stage) is supplied to user B 
at the commitment stage. Since the part of the signal available to user B 
until the disclosure stage should have the same appearance to user B for
both values 0 and 1 of the secret bit (otherwise the user B will have non-zero 
information about the secret bit before the beginning of the disclosure stage),
the part of the classical object which is left with user A should be different 
for different secret bit values. The laws of classical non-relativistic
physics do not prohibit an instantaneous modification of the part of the 
signal still controlled by user A converting 0 into 1 or {\it vice versa}
before submitting it to user B immediately before the disclosure stage
thus allowing cheating by user A. therefore, no unconditionally secure
bit commitment protocol can be realized within the framework of 
non-relativistic classical physics.

In the non-relativistic quantum protocols the information is carried by 
quantum systems. Schematically, the protocols can be described in the 
following way. First, the Hilbert state space ${\cal H}_s$ is chosen to 
which the states of the information carriers belong. User A choses the 
states $|\psi_{0,1}\rangle\in{\cal H}_s$, corresponding to 0 or 1 and  
send them to B. The states are usually chosen to be non-orthogonal. 
It is important that the state space ${\cal H}_s$ is implicitly assumed
to be fully accessible to both users A and B throughout the entire protocol.
The requirement that the density matrices corresponding to both 0 and 1
look identically for user B until the disclosure stage begins results in the
possibility of an undetectable cheating by user A employing an EPR-attack
[11,12]. Roughly speaking, in this approach the protocol involves only the 
state space of the quantum system. However, this situation actually does not
correspond to the real process of information transfer. To be more precise,
we mean the following. The participants of the protocol cannot control the 
entire space. Instead, they control only certain domains (vicinities of their 
laboratories, measuring devices, etc). In addition, all the measurements occur
in the real space and time (or space-time in the relativistic case). The 
non-relativistic quantum mechanics allows construction of entangled states
of physically different systems (we are only interested in this case because 
it is impossible to perform a measurement which affects only one of the two
identical systems). Therefore, if the users control only the non-overlapping
domains, the entangled state from ${\cal H}_s\otimes{\cal H}_a$ should 
automatically be non-local also in the coordinate space. The wave functions
of both systems from ${\cal H}_s$ and ${\cal H}_a$ should be simultaneously
different from zero in the domains controlled by users A and B (because 
otherwise the state will not be entangled for the users). The latter means 
that each user has access to both state spaces, ${\cal H}_s$ and ${\cal H}_a$,
and can perform measurements and unitary transformations separately over 
both systems at his own discretion due to their physical distinguishability.
Therefore, the locality of the transformations in the state space
${\cal H}_s\otimes{\cal H}_a$ (in the sense of manipulation in only one 
of the state subspaces) does not imply the locality in the coordinate 
(position) space. In other words, in the non-relativistic quantum protocols 
of that kind (when the spatio-temporal structure of the information carrier 
states is not explicitly taken into account) the state space of the information
carriers is accessible by both users. In this sense such protocols do not
realize the idea of submitting only a part of information on the carrier 
of a secret bit.

In the non-relativistic case, explicit accounting for the effects of state 
propagation in the position space, when user B has access to only a part of a 
spatially extended state, can hardly introduce any new aspects to the indicated
problems because of the absence of the maximum propagation speed.

Formulation of the problem where only the properties of the state space
${\cal H}$ are used does not correspond to the actual process of information 
transfer in the real space-time. It is more natural to consider a problem
when the users are located in their respective laboratories and control
some their spatial neighborhoods. It is natural to assume that neither A or B
can control and have access simultaneously the entire space.

The state propagation effects (accounting for the spatio-temporal structure 
of the quantum states) were first explicitly used in quantum cryptography
in Ref. [13] (which, in our opinion, was not assessed correctly [14,15]).
Accounting for the restrictions imposed by the special relativity and
quantum mechanics (quantum field theory) [16] substantially simplifies
the proof of unconditional security of relativistic quantum cryptosystems [17].
Besides, the quantum field theory introduces additional fundamental
restrictions, e.g. on the teleportation of quantum states [18].

Recently, the classical bit commitment and distant coin tossing protocols
have been proposed which take into account the existence of finite maximum 
speed of signal (information) propagation [19]. The relativistic classical
protocol [19] is unconditionally secure (i.e. its security is based on the 
fundamental laws of nature only) and in principle allows to delay the second 
stage of the protocol (disclosure of the secret bit value chosen by A) for 
arbitrarily long time. The implementation of this protocol requires that each 
of the users A and B control two spatially separated sites.

The idea of using orthogonal states in the bit commitment and coin tossing
protocols was proposed earlier in Ref. [20]. The protocols suggested in Ref. 
[20] were based on two simple considerations. First, a pair of orthogonal (and,
consequently, reliably distinguishable when completely accessible) 
states become efficiently non-orthogonal (only partly distinguishable) when
restricted to a subspace. This is also true in non-relativistic quantum 
mechanics. Indeed, if we have a pair of spatially extended orthogonal states,
$\psi_{0,1}(x)\in{\cal L}^2(-\infty,\infty,dx)$
\begin{displaymath}
(\psi_0,\psi_1)=\int_{-\infty}^{\infty} \psi_0^*(x)\psi_1(x) dx=0,
\end{displaymath}
they become effectively non-orthogonal when restricted to a subspace 
(a finite domain $\Omega$ in the position space): 
\begin{displaymath} 
(\psi_0,\psi_1)_{\Omega}=\int_{\Omega} \psi_0^*(x)\psi_1(x) dx
\neq 0. 
\end{displaymath}
The second important consideration is the existence of a finite maximum
speed of both quantum states propagation and classical objects motion
implied by the special relativity. This fact does not allow to instantaneously
access the entire state (i.e., the domain where the state is present). 

In contrast to Ref.[20], where the state had no ``internal'' degrees of 
freedom, taking into account the states with ``internal'' degrees of freedom
(e.g. helicity for photons) allows to substantially simplify the protocols.

To be more precise, the states for  
$|\psi_{0,1}\rangle=\psi(x)\otimes|e_{0,1}\rangle$ ($\langle e_0|e_1\rangle=0$)
0 and 1 are orthogonal (due to the internal degrees of freedom) even
if they only partly accessible for the measurement (i.e. the measuring 
apparatus can access only a part of the entire spatial domain where 
$\psi(x)\neq 0$); however, in that case they are not reliably distinguishable:
the probability of distinguishing between these two states (probability
of obtaining a measurement result in one of the orthogonal channels)
can be made arbitrarily small if only a part of the state is accessible
since the probability of obtaining an outcome in a finite spatial domain is
\begin{displaymath}
\langle \psi_i|\psi_i\rangle_{\Omega}=\int_{\Omega}|\psi(x)|^2dx<1, 
\quad i=0,1
\end{displaymath}
so that by appropriately choosing the domain size and function $\psi(x)$ 
the probability can be made arbitrarily small (which actually follows from
the normalization condition $\int_{-\infty}^{\infty} |\psi(x)|^2dx=1$).

Schematically, our protocol can be described in the following way.
User A controls a finite spatial domain and prepares a quantum state in it
at the moment specified by the protocol. This state propagates into the 
quantum communication channel and becomes gradually accessible to the user B
in the spatial domain which is not controlled by user B. Having access to only 
a part of the quantum state in the real position space, user B cannot reliably
determine the secret bit (reliably distinguish between 0 and 1). Moreover,
it is possible to choose the states in such a way that the probability of 
correct determination of the secret bit by user B is arbitrarily close to 1/2
(i.e. simple guessing probability) for arbitrarily long (although agreed upon 
before the start of the protocol) time interval. There exist no fundamental
restrictions on the length of this interval, although making it long enough 
may present a difficult technical problem. The existence of a finite maximum
propagation speed allows to choose the states in such a way that the user A
can no longer modify the chosen bit value after the prepared state has
partly left the domain controlled by him. After the protocol duration time
elapses and the states sent by user A become completely accessible to user B
the latter acquires reliable information on the states with the probability 
arbitrarily close to 1. Important for this protocol are both the quantum nature
of the states involved and the existence of a finite maximum speed of 
propagation imposed by the special relativity.

The states and measurements used in the protocols are described in Section 2
while the bit commitment and coin tossing protocols themselves for the states 
with finite supports are presented in Sections 3 and 4, respectively.
The fundamental non-localizability of quantum states in the quantum field 
theory is accounted for in Section 5. The main results obtained in the paper 
are summarized in Conclusions.

\section{States and measurements used in the protocol.}

Since the protocol explicitly employs the spatio-temporal structure 
of the states, it cannot be formulated without specifying the system geometry.
We shall consider a one-dimensional model containing all the important 
features dictated by the quantum field theory; a similar model is frequently
used in quantum optics. We shall deal with a massless field whose states in 
the momentum representation are specified on the mass shell
$k^2_0-k^2=0$. Important for us are the states propagating in the positive
direction of $x$-axis ($k>0$). We assume that user A controls
a neighbourhood of point $x_A$, while user B controls a neighborhood of 
point $x_B$ ($x_A<x_B$).

In the following all functions are assumed to depend on the difference
$\tau=t-x$; the speed of light is assumed to be unit, 
$c=1$. This representation reflects the intuitive picture of a packet
moving with the speed of light. 
The momentum eigenstate $|k\rangle$, corresponding to the eigenvalue 
$k$ is a generalized eigenvector (to be more precise, a linear continuous
functional on the elements from a dense subset in 
${\cal H}={\cal L}^2(0,\infty,d\xi)$) and has the form
\begin{equation}
\langle \xi |k\rangle=\delta(k-\xi). 
\end{equation}
The states $|\psi\rangle$ from ${\cal H}$ can be expanded in the generalized 
states 
\begin{equation}
|\psi\rangle = \int_{0}^{\infty} \langle k|\psi\rangle |k\rangle dk,
\end{equation}
where the value taken by the functional $\langle k|$ on the element 
$|\psi\rangle$ is
\begin{displaymath}
\langle k| \psi\rangle=\int_{0}^{\infty} \psi(\xi) \delta(k-\xi) d\xi=\psi(k),
\end{displaymath}
i.e. the amplitude of the state $|\psi\rangle$ in the $k$-representation. 
Accordingly, the amplitude of the state $|k\rangle$ in the 
$\tau$-representation is
\begin{equation}
\langle k|\tau\rangle=
\frac{1}{\sqrt{2\pi}}
\mbox{e}^{ik\tau},
\quad k\in (0, \infty), 
\quad 
\tau\in (-\infty,\infty), \tau=t-x,
\end{equation}
corresponding to the intuitive picture of a plane wave (a state with
a definite momentum) moving with the speed of light.

It will be important for the protocol that the quantum states propagate
with the maximum possible speed of light. In the $\tau$-representation  
the orthogonal states (packets) corresponding to 0 and 1 used in the protocol
are written in the form
\begin{equation} 
|\psi_{0,1}\rangle=\int_{-\infty}^{\infty} f(\tau)|\tau\rangle d\tau 
\otimes |e_{0,1}\rangle, 
\quad \
\langle e_0|e_1\rangle=0,
\quad
\langle e_{0,1}|e_{0,1}\rangle=1,
\end{equation}
where the states $|e_{0,1}\rangle$ describe the internal degrees of freedom
(e.g. helicity for photons).

The state normalization condition takes the form
\begin{equation} 
\langle\psi_{0,1}|\psi_{0,1}\rangle=
\int_{-\infty}^{\infty}\int_{-\infty}^{\infty}
f(\tau)f^*(\tau')\langle\tau|\tau'\rangle d\tau d\tau'=1,
\end{equation}
where
\begin{equation} 
\langle\tau|\tau'\rangle=\frac{1}{2\pi}
\delta_+(\tau-\tau')=\frac{1}{2\pi}\int_{0}^{\infty}
\mbox{e}^{ik(\tau-\tau')}dk=
\frac{1}{2}\delta(\tau-\tau')+\frac{i}{\pi}\frac{1}{\tau-\tau'}. 
\end{equation}
We shall introduce the state amplitude in the $k$-representation defined as
\begin{equation}
f(\tau)=\int_{0}^{\infty} f(k)\mbox{e}^{-ik\tau} dk.
\end{equation}
Then taking into account Eqs. (5--7) the normalization condition becomes
\begin{equation} 
\langle\psi_{0,1}|\psi_{0,1}\rangle=
\int_{-\infty}^{\infty}\int_{-\infty}^{\infty}
f(\tau)f^*(\tau')[\frac{1}{2}\delta(\tau-\tau')+
\frac{i}{\pi}\frac{1}{\tau-\tau'}]
d\tau d\tau'.
\end{equation}
Substitution of the state amplitude in the $k$-representation from Eq.(8) 
into Eq.(7) taking into account that [21]
\begin{equation}
\int_{-\infty}^{\infty}
\mbox{e}^{ik\tau} \frac{1}{\tau+a} d\tau=
i\pi\cdot \mbox{sgn}{k}\cdot \mbox{e}^{-iak},
\end{equation}
yields
\begin{equation} 
\langle\psi_{0,1}|\psi_{0,1}\rangle=
\int_{-\infty}^{\infty} |f(\tau)|^2 d\tau = 1.
\end{equation}

The microcausality requirement [22] implies that the field operators
generating the field states belonging to the Hilbert state space when 
acting on the vacuum vector should either commute or anticommute if they
are related to two spatially-like domains. The commutator of two field 
operators is known to be a distribution (see details in Ref. [22]).
If one wishes to talk about local properties of the distribution, the  
test functions should possess certain properties (actually, they 
should belong to the space ${\cal J}(\hat{x})$ of infinitely smooth
functions which vanish at infinity faster than any inverse polynomial).
In other words, the states of a free field cannot have a finite support
(i.e. to be zero outside a finite domain) which means that 
the states of a free field are fundamentally unlocalizable. However,
one construct the states which are arbitrarily strongly localized in space
and vanish at infinity with the rate arbitrarily close to the exponential one 
(e.g. see Refs. [23--27]). In addition, the functions from 
${\cal D}(\hat{x})$ with finite support form a dense set in 
${\cal J}(\hat{x})$ which means that any function from ${\cal J}(\hat{x})$ 
can be approximated with functions from ${\cal D}(\hat{x})$ with any desirable 
accuracy.

In the context of our one-dimensional model the non-localizability can be
derived from the Wiener-Paley theorem [28] since the normalization condition
(10) together with Eq.(7) means the square integrability of the amplitude
in the $k$-representation and imposes the restrictions on the asymptotic
behaviour of the function $f(\tau)$: 
\begin{equation}
f(\tau)=\int_{0}^{\infty} f(k)\mbox{e}^{-ik\tau}dk,
\quad
\int_{-\infty}^{\infty}\frac{\textstyle |\mbox{ln}|f(\tau)||}
{\textstyle 1+\tau^2}d\tau <\infty.
\end{equation}
Equation (11) implies that the function $f(\tau)$ cannot have a finite
support in $\tau$ and cannot decay exponentially at infinity, although
it can be arbitrarily strongly localized and possess decay rate arbitrarily 
close to the exponential law, for example 
\begin{equation}
f(\tau)\propto \exp{\{-\alpha\tau/\mbox{ln(ln...ln} \tau)\}},
\end{equation}
where $\alpha$ take any value.

For the reasons of convenience we shall first formulate the protocol for
the states with finite support (since the functions from ${\cal D}(\tau)$ 
form a dense subset and any function $f(\tau)$ can be approximated with 
functions from ${\cal D}(\tau)$ with any accuracy) and then introduce the 
necessary modification to account for the non-localizability of the states.

Let the state $f(\tau)$ have a finite support, 
supp$f(\tau)=(-\Delta\tau,\Delta\tau)$ ($\Delta\tau$ can be chosen to be 
arbitrarily small). The states are formed by only the vectors 
$|\tau\rangle$ belonging to the interval $(-\Delta\tau,\Delta\tau)$ 
on the light cone:
\begin{equation} 
|\psi_{0,1}\rangle=\int_{-\Delta\tau}^{\Delta\tau}f(\tau)|\tau\rangle 
d\tau\otimes |e_{0,1}\rangle.
\end{equation} 
In contrast to the non-relativistic quantum protocols which do not 
explicitly employ the spatio-temporal structure of the states and the 
state preparation effects are unimportant (to be more precise, the 
non-relativistic quantum mechanics allows an instantaneous preparation 
of any states from ${\cal H}$, even those which are non-local in the position 
space, at any moment of time), the situation is quite different in the field
theory. Preparation of a state requires access to a finite spatio-temporal 
domain (even if the state support is assumed to be finite). In the 
one-dimensional model the state preparation requires either the access to the 
spatial domain of size $\Delta x=2\Delta\tau$ if the state is prepared
by a non-local source at a specified moment of time $t$ or a finite time 
interval $\Delta t=2\Delta\tau/c$ if the state is generated by point-like 
source at point $x$. Therefore in the relativistic case the protocol can only
be formulated after the system geometry is completely specified. The 
one-dimensional situation is the simplest one, since all the quantities here 
depend on a single variable $\tau=x-ct$. Bearing in mind that the actual 
experiments employ quasi-one-dimensional optical fiber systems, analysis of the
one-dimensional model seems to be quite reasonable.

Consider now the ``stretched'' states used in the protocols. These states 
consist of the two halves separated by the interval $\tau_0$ on the light cone
and can be written as
\begin{equation} 
|\psi_{0,1}(\tau_0)\rangle=
\frac{1}{\sqrt{2}}\int ( f(\tau)+ f(\tau-\tau_0) )|\tau\rangle)
d\tau\otimes |e_{0,1}\rangle.
\end{equation}
Here and below we adopt the normalization
\begin{equation}
\int_{-\infty}^{\infty}|f(\tau)|^2 d\tau= 
\int_{-\infty}^{\infty} |f(\tau-\tau_0)|^2 d\tau=1,
\quad
\mbox{supp}f(\tau)\cap \mbox{supp}f(\tau-\tau_0)=\emptyset.
\end{equation}

Since the initial state support belongs to the interval 
$(-\Delta\tau,\Delta\tau)$, the state preparation requires
access to the domain $(-\Delta\tau,\Delta\tau+\tau_0)$ on the light cone
(either domain $\Delta x=(-\Delta\tau,\Delta\tau+\tau_0)$ of the position 
space if the preparation is performed with a non-local apparatus at a 
specified moment of time or the time interval
$\Delta t=(-\Delta\tau,\Delta\tau+\tau_0)$ if the state is produced by 
a local source at point $x$).

Consider now the individual measurements performed by user B over the
quantum field states. The measurements are described by a partition 
of unity specified on the possible outcomes space defined as a set
$\Omega=\{\tau\in(-\infty,\infty),i=0,1\}$:
\begin{equation} 
I=\left(\int_{-\infty}^{\infty}{\cal M}(d\tau)\right)\otimes
\left({\cal P}_0+{\cal P}_1 \right)=
\end{equation}
\begin{displaymath}
\left(\int_{-\infty}^{\infty}d\tau
\left( \int_{0}^{\infty} \mbox{e}^{ik\tau}|k\rangle dk\right)
\left( \int_{0}^{\infty} \mbox{e}^{-ik'\tau}\langle k'| dk'\right)\right)
\otimes\left({\cal P}_0+{\cal P}_1 \right),
\quad
\langle k|k'\rangle =\delta(k-k')
\end{displaymath}
where $|k\rangle$ is a formal eigenvector with the specified $k$, and
\begin{displaymath}
{\cal M}(d\tau)=
|\tau\rangle\langle \tau| d\tau,
\quad
{\cal P}_{0}=|e_0\rangle\langle e_0|,\quad
{\cal P}_{1}=|e_1\rangle\langle e_1|.
\end{displaymath}
We shall also need the description of the state propagation through the
quantum communication channel from the domain controlled by user A to the 
domain controlled by user B. This propagation is described by a unitary 
translation of the state $|\psi_{0,1}\rangle$ along the light cone branch
$\tau=x-ct$:
\begin{equation} 
{\cal U}_{ch}(\tau_{ch})|\psi_{0,1}\rangle=
|\psi_{0,1,\tau_{ch}}\rangle=
\frac{1}{\sqrt{2}}
\int ( f(\tau-\tau_{ch}) + f(\tau-\tau_0-\tau_{ch}) )\rangle)d\tau
\otimes|e_{0,1}\rangle;
\end{equation}
here $\tau_{ch}$ is the channel length. The state ``extent''
($2\Delta\tau+\tau_0$) and the communication channel length ($\tau_{ch}$) 
should satisfy the inequality $\tau_{ch}<\tau_0+2\Delta\tau$, so that
one can assume without loss of generality that $\tau_{ch}=0$  
(the channel length can be arbitrary until it does not exceed
the state ``extent'').

The probability of obtaining an outcome by user B in the channel
$i$ (${\cal P}_i$) in the interval $d\tau$ for the input state 
$|\psi_j(\tau_0)\rangle$ is
\begin{equation} 
\mbox{Pr}\{d\tau;i,j\}=\mbox{Tr}
\left\{
\left(\left({\cal M}(d\tau)\right)\otimes{\cal P}_i\right)
|\psi_j(\tau_0)\rangle\langle\psi_j(\tau_0)|
\right\}=
\delta_{ij}\frac{1}{2}\left\{|f(\tau)|^2+|f(\tau-\tau_0)|^2\right\} d\tau.
\end{equation}
This expression describes the probability density for obtaining an outcome 
in one of the orthogonal (distinguishable) channels for 0 ($i=j=0$) and 1 
($i=j=1$) in the interval $d\tau$. At the intuitive level such a measurement
can be thought of as being realized with a very fast (formally with zero
intrinsic time) photodetector operating in waiting mode. The measurement 
outcome is a random event occurring in the time interval $d\tau$ with the
probability density given by Eq.(18).

The probability of detecting a state in the finite interval $\Delta(\tau)$
(for $i=j$) is
\begin{equation} 
\mbox{Pr}\{\Delta(\tau)\}=\int_{\Delta(\tau)}\mbox{Pr}\{d\tau;i,i\}=
\frac{1}{2}\left\{
\int_{\Delta(\tau)}|f(\tau)|^2d\tau+
\int_{\Delta(\tau)}|f(\tau-\tau_0)|^2d\tau\right\}.
\end{equation} 
If the interval $\Delta(\tau)$ (accessible domain on the light cone) does not
entirely cover the state support (for example, if only one half of the state 
is covered), the probability of obtaining an outcome is 1/2. 
However, if an outcome is obtained the states are uniquely identified
because of the orthogonality of the channels ${\cal P}_0$ and ${\cal P}_1$.
Therefore, in the time interval $\Delta\tau\le \tau\le\tau_0+\Delta\tau$ 
the probability of wrong state identification based on the measurement outcome 
is 1/4. Accordingly, the probability of correct identification is 3/4.
On the other hand, for simple guessing the error probability is 1/2.

It should be noted once again that the above measurement cannot be
interpreted as a measurement which lasts for a finite time $\Delta(\tau)$: 
Every outcome occurs randomly at time $t$ with the probability density (18).

After the time $\tau_0+2\Delta\tau$ elapses and the entire state is found
in the domain controlled by user B, the values 0 and 1 are uniquely identified
because of the orthogonality of the corresponding states.

Hence, propagation of the states with the maximum possible speed allows 
an explicit and natural implementation of the idea of providing by user A of
only a part of information (part of a quantum state) on the chosen secret bit.
Quantum nature of the state is important for the protocol since for a classical 
signal whose shape is described by the function $f(\tau)$ 
with different polarizations $e_0$ or $e_1$ the probability of correct 
identification is 1 (rather than 3/4) even if only a part of the signal 
is accessible. The correct identification probability of 3/4 in the quantum 
case is actually a consequence of the state normalization requirement.

This result can also be derived in a somewhat different way allowing to 
clarify the peculiarity of the situations where only a part of the Hilbert 
state is accessible for measurements. Let us find the measurement minimizing
the identification error in the problem of distinguishing between the two 
density matrices where the states are only partly accessible. 
The density matrices are written in the form
\begin{equation}
\rho_{0,1}=\left\{
\frac{1}{\sqrt{2}}
\left(
\int_{-\infty}^{\infty} [f(\tau)+f(\tau-\tau_0)]|\tau\rangle d\tau
\right)
\left(
\int_{-\infty}^{\infty} [f^*(\tau')+f^*(\tau'-\tau_0)]\langle\tau'| d\tau'
\right)
\right\}
\otimes
|e_{0,1}\rangle\langle e_{0,1}|=
\end{equation}
\begin{displaymath}
\rho(f)\otimes\rho(0,1).\makebox[8cm]{ \ \ }
\end{displaymath}
We shall next derive an expression for the identification error occurring
when trying to distinguish between the states $\rho_{0,1}$ under the conditions
where only a part of the entire space-time is accessible for measurements.
Formally, the problem is reduced to the case where the domain $\Delta(\tau)$
is accessible for the measurements while the rest of the space-time
(denoted as $\overline{\Delta}(\tau)=(-\infty,\infty)- \Delta(\tau)$) cannot
be accessed by available measuring apparatus. 

The measurement is described by the resolution of identity consisting of 
two terms. The first one is actually the identity operator in the subspace 
spanned by the basis vectors $|\tau\rangle$ belonging to the interval
$\Delta(\tau)$, while the second term is the identity operator in the subspace 
spanned by the vectors form the inaccessible domain
$\overline{\Delta}(\tau)=(-\infty,\infty)- \Delta(\tau)$:  
\begin{equation} 
I\otimes {\bf C}^2=
I(\Delta\tau)\otimes {\bf C}^2+
I(\overline{\Delta}(\tau))\otimes {\bf C}^2=
\left(\int_{\Delta(\tau)} |\tau\rangle\langle \tau| d\tau\right)\otimes 
{\bf C}^2+
\left(\int_{\overline{\Delta}(\tau)} |\tau\rangle\langle \tau| d\tau\right)
\otimes 
{\bf C}^2.
\end{equation} 
Suppose that the state $\rho_0$ is produced for the measurements with the 
{\it a priori} probability $\pi_0$, while the state $\rho_1$ with the 
probability $\pi_1$ ($\pi_0+\pi_1=1$). In the following we shall assume 
that $\pi_0=\pi_1=1/2$, i.e. 0 or 1 are chosen by the user A with equal 
probabilities. 

Since only a part of the space-time is accessible for measurements (which 
automatically implies the restricted access to the Hilbert state space
since the basis states are labeled by $\tau$), the total error contains 
two terms. The first one ($P_e(\overline{\Delta}(\tau))$) corresponds to the
situation where the measuring apparatus (photodetector) operated by user B
did not fire (the outcome occurred in the inaccessible domain). The second term
($P_e(\Delta(\tau))$) describes the error in the state identification arising 
in the case where the measurement outcome took place in the domain accessible 
for user B.

The probability of the event when the state was not detected by user B
(his measuring apparatus did not fire) is
\begin{equation}
P(\overline{\Delta}(\tau))=\mbox{Tr}
\{ (\pi_0\rho_0+\pi_1\rho_1) 
\left(I(\overline{\Delta}(\tau))\otimes {\bf C}^2 \right)
\} = \pi_0  p_0 +\pi_1 p_1, \quad p_0=p_1=p,
\end{equation}
\begin{displaymath}
p=\frac{1}{2} \int_{\overline{\Delta}(\tau)} 
[|f(\tau)|^2+|f(\tau-\tau_0)|^2]d\tau.
\end{displaymath}
The probability $p_0$ of the outcome in the inaccessible domain for 
the input state $\rho_0$ produced with the specified {\it a priori} 
probability $\pi_0$ is
\begin{equation}
p_0=\frac{\textstyle \pi_0 p}{\textstyle \pi_0 p + \pi_1 p}= \pi_0,
\end{equation}
and, similarly, for $\rho_1$ produced with the {\it a priori} probability
$\pi_1$,
\begin{equation}
p_1=\frac{\textstyle \pi_1 p}{\textstyle \pi_0 p + \pi_1 p} = \pi_1.
\end{equation}
The probability of error, i.e. the probability of the event when 
the state $\rho_1$ is interpreted as state $\rho_0$, and {\it vice versa}, is
\begin{equation}
P_e(\overline{\Delta}(\tau))=\pi_0 p_1+\pi_1 p_0.
\end{equation}
If only one half of the state is located in the accessible domain
(the support of either $f(\tau)$ or $f(\tau-\tau_0)$ belongs to the
inaccessible domain), the probability of wrong identification
for the case of the outcome occurring in the inaccessible domain 
calculated according to Eqs. (22--25) is $P_e(\overline{\Delta}(\tau))=1/2$.

In the general case the measurement minimizing the wrong identification 
probability is given for a binary resolving function by the identity resolution
\begin{equation} 
\tilde{E}_0+\tilde{E}_1=I(\Delta(\tau))\otimes I_{{\bf C}^2}=
I(\Delta(\tau))\otimes \left( E_0+E_1 \right),
\end{equation} 
where, in contrast to Refs. [29,30], the resolution is specified in the
subspace restricted to $\Delta(\tau)$. 
The minimal error probability is found by the optimization with respect to all 
possible resolutions (see details in Refs. [29,30]):
\begin{equation} 
P_e(\Delta(\tau))=\min_{ \{\tilde{E}_0,\tilde{E}_1\}  }
\left(
\pi_0 \mbox{Tr}\{\rho_0\tilde{E}_1\}
+
\pi_1 \mbox{Tr}\{\rho_1\tilde{E}_0\}
\right),
\end{equation} 
where $\pi_0$ and $\pi_1$ are the probabilities of occurrence of density 
matrices $\rho_0$ and $\rho_1$, respectively (in our problem $\pi_0=\pi_1=1/2$ 
are the probabilities of the preparation of 0 and 1 by user A).

Taking into account Eq.(26), the error probability can be reduced to the 
following form:
\begin{equation} 
P_e(\Delta(\tau))=\pi_0\mbox{Tr}\{ \rho(f) I(\Delta(\tau)) \}
+
\mbox{Tr}\{ \Gamma\tilde{E}_0\},
\end{equation} 
\begin{displaymath}
\Gamma=\pi_1\rho_1-\pi_0\rho_0.
\end{displaymath}
Optimization of $P_e(\Delta(\tau))$ reduces to finding the minimum 
of $\mbox{Tr}\{\Gamma\tilde{E}_0\}$ with respect to all possible operators
$\tilde{E}_0$. Since the domain accessible for measurements is restricted to
the interval $\Delta(\tau)$, one has
\begin{equation} 
\mbox{Tr}\{ \Gamma\tilde{E}_0 \}=
\mbox{Tr}_{ \Delta(\tau) } 
\{ \Gamma \tilde{E}_0 \}=
\mbox{Tr}_{ \Delta(\tau)  } 
\{ 
\rho(f)\}\otimes \mbox{Tr} 
\{
(\pi_1\rho(1)-\pi_0\rho(0)) E_0 
\}=
\end{equation} 
\begin{displaymath}
\left\{ 
\frac{1}{2} \int_{ \Delta(\tau) } [|f(\tau)|^2+|f(\tau-\tau_0)|^2]d\tau
\right\}
\otimes \mbox{Tr}\{\Gamma E_0\}.
\end{displaymath}
Since $0\le E_0\le I(\Delta(\tau))\otimes I_{{\bf C}^2}$,  
\begin{equation}
\mbox{Tr}\{\Gamma E_0\}\ge
\mbox{Tr}\{\Gamma \}=\sum_i \gamma_i
\end{equation}
The minimal possible error is determined by the negative eigenvalues
$\gamma_i$ of the operator $\Gamma=\pi_1\rho(1)-\pi_0\rho(0)$ [30]. The operator
$\tilde{E}_0$ should satisfy the conditions
\begin{equation} 
\langle \gamma_i|\tilde{E}_0|\gamma_i\rangle=1, \quad \gamma_i\le 0,
\end{equation} 
\begin{displaymath}
\langle \gamma_i|\tilde{E}_0|\gamma_i\rangle=0, \quad \gamma_i\ge 0,
\end{displaymath}
where $|\gamma_i\rangle$ are the eigenvectors of the operator 
$\Gamma=\sum_i \gamma_i |\gamma_i\rangle\langle \gamma_i|$.
In the basis $\{ |e_0\rangle,|e_1\rangle \}$ the operator $\Gamma$ 
has the matrix
\begin{equation} 
\Gamma=
\left(
\begin{array}{cc}
 \pi_1 & 0 \\
   0   & -\pi_0
\end{array}
\right),
\quad \mbox{negative}\quad \gamma_2=-\pi_0=-1/2.
\end{equation} 
Accordingly, the operator $\tilde{E}_0$ is
\begin{equation}
\tilde{E}_0=I(\Delta(\tau))\otimes 
\left(
\begin{array}{cc}
   0   &  0 \\
   0   &  1
\end{array}
\right),
\quad
\tilde{E}_1=I(\Delta(\tau))\otimes 
\left(
\begin{array}{cc}
   1   &  0 \\
   0   &  0
\end{array}
\right).
\end{equation}
For the minimal error one obtains 
\begin{equation} 
P_e(\Delta(\tau))= 
\overline{f^2(\Delta(\tau))} (\pi_0+\sum_{\gamma_i\le 0}\gamma_i),
\end{equation} 
where the notation
\begin{equation} 
\overline{f^2(\Delta(\tau))}=
\frac{1}{2} \int_{ \Delta(\tau) } [|f(\tau)|^2+|f(\tau-\tau_0)|^2]d\tau.
\end{equation} 
is introduced. Finally, for the error probability in the case where an outcome
occurred in the domain accessible for measurements one has
\begin{equation} 
P_e(\Delta(\tau))= \overline{f^2(\Delta(\tau))}(\pi_0-\pi_0)=
\frac{1}{2}(\frac{1}{2}-\frac{1}{2})=0,
\end{equation} 
\begin{displaymath}
\overline{f^2(\Delta(\tau))}=\frac{1}{2}.
\end{displaymath}
At the intuitive level this result can be interpreted in the following way.
Suppose that one has to distinguish between a pair of single-photon 
extended states with different (orthogonal) helicities. The correct
identification probability is only unit if the entire states are accessible 
for the measurements (accordingly, the error probability is zero). In spite of
the orthogonality of the basis vectors describing different helicities, the
states cannot be reliably identified due to their spatial extent if they
(their spatial amplitudes) are not entirely accessible. Physically, this is 
related to the fact that there exist no helicity states beyond the spatial 
degrees of freedom. Because of the normalization condition with respect to the
spatial degrees of freedom the probability of firing of any measuring device 
employed by user B does not exceed 1. Reliable distinguishing of the two states 
with even orthogonal helicities always requires a finite time since because of 
the restrictions imposed by special relativity the entire state cannot be 
accessed faster than the effective state ``extent'' divided by the speed of 
light.

For a large number of measurements the total error is the relative frequency of
wrongly identified states. The fraction (probability) of outcomes in the 
accessible domain is
\begin{equation}
N(\Delta(\tau))=\mbox{Tr}\{
(\pi_0\rho_0+\pi_1\rho_1) (I(\Delta(\tau))\otimes I_{{\bf C}^2})
\},
\end{equation}
and, similarly, the fraction of outcomes in the inaccessible domain is
\begin{equation}
N(\overline{\Delta}(\tau))=\mbox{Tr}\{
(\pi_0\rho_0+\pi_1\rho_1) (I(\overline{\Delta}(\tau))\otimes I_{{\bf C}^2})
\}.
\end{equation}
The total error probability $P_e$ is the sum of the error occurring for
the outcome taking place in the inaccessible domain multiplied by the 
probability of these outcomes and the product of the error occurring for
the outcome taking place in the accessible domain and the 
probability of these outcomes: 
\begin{equation}
P_e=P_e(\overline{\Delta}(\tau))\cdot N(\overline{\Delta}(\tau)) + 
P_e(\Delta(\tau))\cdot N(\Delta(\tau)).
\end{equation}
When only halves of the states are accessible, one has
\begin{equation}
P_e=\frac{1}{2}\cdot\frac{1}{2} + 0\cdot\frac{1}{2},
\end{equation}
\begin{displaymath}
P_e(\overline{\Delta}(\tau))=\frac{1}{2},
\quad
N(\overline{\Delta}(\tau))=\frac{1}{2},
\quad
P_e(\Delta(\tau))=0,
\quad
N(\Delta(\tau))=\frac{1}{2}.
\end{displaymath}
Accordingly, the correct identification probability is 3/4.

The obtained result actually means the following. If user A prepares randomly 
and with equal probabilities either the state $\rho_0$ or $\rho_1$ and
sends these states for measurements to user B, the probability of firing of 
the measuring device in one of the channels corresponding to 0 or 1
is 1/2. If the measuring apparatus employed by user B gave an outcome, 
the state is reliably identified. However, if the apparatus did not fire,
the user B can only guess which state was sent to him by user A.
The probability for the apparatus not to fire is 1/2; in that case
the correct guess probability is also 1/2. Hence for these events the
net probability of correct identification is $1/2\cdot 1/2=1/4$. The total
probability of correct identification of the state sent by user A is thus
$1/2+1/4=3/4$.

In this case the probability $1-P_e$ coincides with the probability of correct 
identification of the secret bit.

This probability of correct identification is too high (substantially exceeds 
1/2) for development of a valid protocol. The situation is radically changed 
if the secret bit is constructed as parity bit of $N$ states. We shall see 
below that in this case the probability of correct identification of the 
parity bit exceeds 1/2 by only an exponentially small value, i.e. practically 
coincides with the probability of simple guessing which is the worst strategy
for user B.

Let us now calculate the probability of the error of identification of 
the secret bit value when it is coded as parity bit of $N$ orthogonal states.
For the case where the entire state space is accessible for the measurements,
the problem of the parity bit of a string of $N$ bits each coded by one
of the two non-orthogonal states was considered earlier in Ref. [32]. 

We shall first calculate the identification error for the outcomes occurring 
in the accessible domain. For a random string of $N$ bits associated with
the density matrix $\rho_{0,1}$ described by $2^N$ possible combinations
($2^N/2$ of which are even and $2^N/2$ are odd) the problem is reduced to 
distinguishing between the two density matrices corresponding to even and 
odd strings:
\begin{equation} 
\hat{\rho}_0=\frac{2}{2^N}\sum_{ (i_1\oplus i_2\oplus \ldots i_N)=0}
\overbrace{\rho_{i_1}\otimes\rho_{i_2}\otimes\ldots\rho_{i_N}  }^{N}=
\end{equation} 
\begin{displaymath}
\frac{2}{2^N}
\left(
\rho(f)\otimes\rho(f)\otimes\ldots\rho(f)
\right)
\otimes
\sum_{ (i_1\oplus i_2\oplus \ldots i_N)=0}
\rho(i_1)\otimes\rho(i_2)\otimes\ldots\rho(i_N),
\quad 
i_k = 0,1,
\quad
k=1,\ldots N;
\end{displaymath}
\begin{equation} 
\hat{\rho}_1=\frac{2}{2^N}\sum_{ (i_1\oplus i_2\oplus \ldots i_N)=1}
\overbrace{\rho_{i_1}\otimes\rho_{i_2}\otimes\ldots\rho_{i_N}  }^{N}=
\end{equation} 
\begin{displaymath}
\frac{2}{2^N}
\left(
\rho(f)\otimes\rho(f)\otimes\ldots\rho(f)
\right)
\otimes
\sum_{ (i_1\oplus i_2\oplus \ldots i_N)=1}
\rho(i_1)\otimes\rho(i_2)\otimes\ldots\rho(i_N),
\quad 
i_k = 0,1
\quad
k=1,\ldots N.
\end{displaymath}
The measurement minimizing the identification error for the density matrices
$\hat{\rho}_0$ and $\hat{\rho}_1$ is given by the identity resolution
\begin{equation} 
\left(I(\Delta(\tau))\otimes I_{{\bf C}^2}\right)^{\otimes^N}=
\tilde{\hat{E}}_0+\tilde{\hat{E}}_1=
I(\Delta(\tau))^{\otimes^N}\otimes 
\left( \hat{E}_0+\hat{E}_1 \right),
\quad
\hat{E}_0+\hat{E}_1=I_{{\bf C}^2}^{\otimes^N}.
\end{equation} 
In that case the error probability is
\begin{equation} 
P_e( \Delta(\tau) )=
\pi_0\mbox{Tr}\left\{\hat{\rho}_0 \left(I(\Delta(\tau))^{\otimes^N}\otimes
I_{{\bf C}^2}^{\otimes^N}\right) \right\}+ 
\mbox{Tr}\left\{ \hat{\Gamma} 
\left( I(\Delta(\tau))^{\otimes^N}\otimes\hat{E}_0 \right)  
\right\},
\end{equation} 
Accordingly, the minimal error is given by a formula similar to Eq.(28), 
and is determined by the negative eigenvalues ($\gamma_i$) of the operator
\begin{equation} 
\hat{\Gamma}=\left(\overline{f^2(\Delta)}\right)^N
\left(\pi_1\hat{\rho}(1)-\pi_0\hat{\rho}(0)\right)=
\left(\overline{f^2(\Delta)}\right)^N \Gamma.
\end{equation} 
In the basis of vectors ordered into the even and odd (with respect to the
sum of subscripts) sets, the operator $\Gamma$ is written as
\begin{equation} 
\Gamma=
\frac{1}{2^N}
\left(
\begin{array}{cccc}
1 & 0 & 0 & \ldots\\
0 & 1 & 0 & \ldots\\
. & . & . & \ldots\\
. & . & . & \ldots\\
0&  \ldots& -1& 0 \\ 
0&  0 & \ldots& -1\\
\end{array}
\right),
\quad
\begin{array}{c}
\mbox{odd}
\left\{
\begin{array}{c}
|e_0\rangle\ldots |e_1\rangle\\
\ldots \\
|e_1\rangle\ldots |e_0\rangle\\
\end{array}
\right.
\\
\mbox{ }\mbox{ }\mbox{even}
\left\{
\begin{array}{c}
|e_0\rangle\ldots |e_0\rangle\\
\ldots \\
|e_1\rangle\ldots |e_1\rangle\\
\end{array}
\right.
\end{array}.
\end{equation} 
Finally, the minimal error probability of the identification of the parity 
bit determined by $N$ orthogonal states which are only partly accessible, is
($\pi_0=1/2$)
\begin{equation} 
P_e( \Delta(\tau) )=\left(\overline{f^2(\Delta)}\right)^N
\left(
\frac{1}{2}+\left(\frac{1}{2^N} \sum_{\gamma_i\le 0}^{2^N/2}(-1) \right)
\right).
\end{equation} 
If only one half of each state is accessible ($\overline{f^2(\Delta)}=1/2$),
one has
\begin{equation} 
P_e( \Delta(\tau) )=\left(\frac{1}{2}\right)^N
\left( \frac{1}{2} - \frac{1}{2^N}\cdot \frac{2^N}{2} \right)=0.
\end{equation} 
If the measurement outcome took place in the accessible domain, the 
identification error is zero because of the channel orthogonality.
This formula should be understood in the following way. If all $N$ outcomes
occurred in the accessible domain, the error probability is zero because of 
the channel orthogonality. The same is true if the outcomes in the accessible 
domain took place for $m$ states (in that case $N$ in Eqs.(47,48) should 
be replaced by $m$). In other words, the states which resulted in 
measurement outcomes occurring in the accessible domain become reliably known
to user B.

However, the outcomes can also occur in the inaccessible domain.

As the states gradually propagate into the domain accessible for the 
measurements performed by user B, $\overline{f^2(\Delta)}\rightarrow 1$, 
the error probability $P_e\rightarrow 0$. Any two orthogonal states are
reliably distinguishable when each of them is entirely accessible for
the measuring apparatus.

Let us now calculate the probability of correct identification of the parity 
bit. The total number of binary strings of length $N$ is $2^N$. The outcomes
can occur both in accessible and inaccessible domains. The total space of 
outcomes can be divided into two disjoint subsets. The first one corresponds to 
the event when all $N$ outcomes occurred in the accessible domain. In that case 
the probability of correct identification of the parity bit is 1. However, 
the probability of this event for the case where only one half of each state 
is accessible $\overline{f^2(\Delta)}^N=2^{-N}$.

The second subset corresponds to all other events
when at least one outcome occurred in the inaccessible domain. The probability 
of all these events (when only one half of each state is accessible) is
\begin{equation}
\sum_{k=0}^{N-1} C^{k}_{N}
\overline{f^2(\Delta)}^{k}(1-\overline{f^2(\Delta)})^{N-k}=
\sum_{k=0}^{N-1} C^{k}_{N}
\frac{1}{2^{k}}\frac{1}{2^{(N-k)}}=1-\overline{f^2(\Delta)}^{N}=
1-2^{-N}.
\end{equation}
For these events the probability of error in the parity bit identification
is 1/2. Indeed, if user B has a string of length $k$ ($k\le N-1$) whose parity
is reliably known to him. However, the parity of the rest part of the string
of length $N-k$ corresponding to the outcomes in the inaccessible domain
can be either odd or even with equal probabilities. Hence the parity of the 
full string consisting of $N$ bits is known with the probability of 1/2,
since the knowledge of the string of $k$ bits does not help in any way in 
finding the parity bit of the full string. 

The total error in the parity bit determination is a sum of two contributions. 
The first one corresponds to the event when all outcomes took place in the
accessible domain and the second one corresponds to all the rest events. Each 
contribution is a product of the probability of error in the parity bit 
identification and the probability of the event itself. One finally has
\begin{equation}
P_e(parity)=\frac{1}{2}\cdot\left(1 - 2^{-N}\right) + 0\cdot 2^{-N}=
\frac{1}{2}- \frac{1}{2}\cdot 2^{-N}.
\end{equation}
Accordingly, as long as only one half of each state is accessible to user B,
the probability of correct parity bit identification by this user is 
\begin{equation}
P_c(parity)=1-P_e(parity)=\frac{1}{2} + \frac{1}{2}\cdot 2^{-N}
\end{equation}
and exceeds the simple guess probability by only an exponentially small value.

Thus, during the time interval $\tau_0$ 
($\Delta\tau <\tau <\Delta\tau+\tau_0$) after the beginning of the protocol
the user B has only exponentially small information on the secret bit.

However, this scheme where the secret bit is coded as a parity bit of a string
of $N$ bit is still insufficient for the development of a useful protocol since
it allows the user A to cheat with unacceptably high probability (to delay
his choice of the secret bit without being caught by user B).

To avoid this problem, each of $N$ bits should be coded by a block of
$k$ identical bits (the number $k$ will be specified below) which
are randomly distributed over $N\cdot k$ channels.

Finally, we shall give an expression for the probability of error when
distinguishing between the two density matrices corresponding to 0 and 1
for the case where the parity bit is coded by the blocks of identical
bits (all of them are either 0 or 1). In the protocol the secret parity bit
is calculated over $N$ bits each of which is represented by a block of
length $k$. This block-wise representation of each bit is necessary for 
the detection by user B of possible cheating by user A.

In that case the total number of binary string is $2^{N\cdot k}$. When each 0 
and 1 is coded by blocks of length $k$, the number of odd and even strings 
among them is
\begin{equation} 
S_{odd,even}=\frac{1}{2}\sum_{m=0}^{N-1}C^{m\cdot k}_{N\cdot k}=
2^{N\cdot k}\left( \frac{1}{2k} \right)
\sum_{l=1}^{k} 
\mbox{cos}^{N\cdot k}\left(\frac{l\pi}{k}\right)
\mbox{cos}\left({Nl\pi}\right)
\approx 2^{N\cdot k},
\end{equation} 
i.e. is actually equal to the number of different ways of distributing
$(N-l)\cdot k$ units and $l\cdot k$ zeros (for $0\le l \le N$) among 
$N\cdot k$ cells [30].

Note that if the position of each block was agreed upon in advance (i.e.
the units and zeros from different blocks were not intermixed) the total
number of possible odd and even strings would only be $2^N$ which is 
exponentially less than $2^{-N\cdot k}$ (52) for large $k$.

It will be important for the protocol that the rest 
$N_{rest}=2^{N\cdot k}-S_{odd}-S_{even}\ll 2^{-N\cdot k}$ strings do not 
belong to either even or odd string sets coded by blocks of length $k$.
 
In the complete basis ordered with respect to odd and even block-wise 
states and other states (for definiteness we assume $k$ to be even)
the operator similar to Eq. (46) is 
\begin{equation} 
S_{even}\rightarrow
\left\{
\begin{array}{ccc}
\overbrace{|e_0\rangle\otimes|e_0\rangle\ldots}^{k} & \ldots
\overbrace{|e_0\rangle\otimes|e_0\rangle\ldots}^{k} \\
\mbox{other permutations of 0 and 1}  & \ldots &  \\
\overbrace{|e_1\rangle\otimes|e_1\rangle\ldots}^{k} & \ldots
\overbrace{|e_1\rangle\otimes|e_1\rangle\ldots}^{k} \\
\end{array}
\right.
\end{equation} 
\begin{equation} 
S_{odd}\rightarrow
\left\{
\begin{array}{ccc}
\overbrace{|e_0\rangle\otimes|e_0\rangle\ldots}^{k} & \ldots
\overbrace{|e_1\rangle\otimes|e_1\rangle\ldots}^{k} \\
\mbox{other permutations of 0 and 1}  & \ldots &  \\
\overbrace{|e_1\rangle\otimes|e_1\rangle\ldots}^{k} & \ldots
\overbrace{|e_0\rangle\otimes|e_0\rangle\ldots}^{k} \\
\end{array}
\right.
\end{equation} 
the operator similar to Eq. (46) takes the form
\begin{equation} 
\hat{\Gamma}=
\left(
\begin{array}{ccc}
\hat{I}_{S_{odd}} &       0      &   0   \\
0           &-\hat{I}_{S_{even} }&   0    \\
0           &       0      &   \hat{0}    \\
\end{array}
\right),
\end{equation} 
where $\hat{I}_{S_{odd}}$ ($\hat{I}_{S_{even}}$) are unit
$S_{odd}\times S_{odd}$ ($S_{even}\times S_{even}$) matrices, and 
$\hat{0}$ is the zero matrix of size $N_{rest}\times N_{rest}$.

The measuring operators $\hat{E}_0$ and $\hat{E}_1$ in the same basis are 
written as
\begin{equation} 
\hat{E}_0=
\left(
\begin{array}{ccc}
\hat{0}     &       0      &   0   \\
0           & \hat{I}_{S_{even} }&   0    \\
0           &       0      &   \hat{0}    \\
\end{array}
\right),
\quad
\hat{E}_1=
\left(
\begin{array}{ccc}
\hat{I}_{S_{odd} } &       0      &   0   \\
0                  &       \hat{0}&   0    \\
0                  &       0      &   \hat{I}    \\
\end{array}
\right).
\end{equation} 

The probability of error in identification of the block-wise parity bit under 
the conditions that $N\cdot k$ outcomes took place in the accessible domain is
\begin{equation} 
P_e(\Delta(\tau))=\left(\overline{f^2(\Delta)}\right)^{N\cdot k}
\left( \frac{1}{2}-\frac{1}{2^{(S_{even}+S_{odd} )}}
\sum_{\gamma_i\le 0}^{2^{S_{even}} }(-1) \right)=0.
\end{equation} 
The error probability is zero for all the states which gave the 
outcomes in the accessible domain. 

Let us now calculate the error in identification of the parity bit when coding
with the blocks of length $k$ is adopted. The outcomes can occur both in the 
accessible and in inaccessible domains. We shall first calculate the minimal 
number of outcomes which should occur in the accessible domain if the string
parity is to be reliably identified. Since the direct calculation is rather 
difficult, we shall take advantage of the following approach (which is actually
a straightforward modification of the Shannon typical sequence method [33,34]).
For a moment we shall return to the situation where each bit is represented by
a block of unit length, $k=1$. Since the set of all possible strings contains  
$\Omega=2^N$ elements, the information carried by each particular string 
is $I=\mbox{log}_2 |\Omega|$ and (to within the rounding error) coincides with
the number of binary symbols required to specify each string. If each symbol
(in our case, firing of the detector employed by user B in the accessible 
domain) occurs with the probability $p$, the probability of the element 
identification is $p^I$.

For block-wise coding the number of all possible strings is given by Eq. (52)
and the number of binary symbols required to identify a particular string is
\begin{equation}
I=\mbox{log}_2 
\left(
\frac{2^{N\cdot k-1}}{k}
\sum_{l=1}^{k} 
\mbox{cos}^{N\cdot k}\left(\frac{l\pi}{k}\right)
\mbox{cos}\left({Nl\pi}\right)
\right)=\alpha(N,k)(N\cdot k),
\end{equation}
which yields the number of outcomes in the accessible domain required 
to identify the string parity.

Accordingly, the probability of this event is ($p=\overline{f^2(\Delta)}=1/2$)
\begin{equation}
P_{acc}=p^{\alpha(N,k)(N\cdot k)}=2^{-\alpha(N,k)(N\cdot k)}.
\end{equation}
For these outcomes the error probability is zero. Accordingly, for the 
outcomes in the inaccessible domain we have  
\begin{equation}
P_{unacc}=1-P_{acc}=1-2^{-\alpha(N,k)(N\cdot k)};
\end{equation}
the parity bit identification error in that case is
\begin{equation}
P_e(parity)=\frac{1}{2}\cdot\left(1-2^{-\alpha(N,k)(N\cdot k)}\right)+
0\cdot 2^{-\alpha(N,k)(N\cdot k)}.
\end{equation}
Hence, the correct parity bit identification probability exceeds the simple 
guess probability by only an exponentially small value:
\begin{equation}
P_c(parity)=1-P_e(parity)=\frac{1}{2} +  2^{-\alpha(N,k)(N\cdot k)}.
\end{equation}

Note that in the order of magnitude the number of block-wise coded
strings with the zeros and units of all blocks randomly distributed over 
the entire string is equal to the total number of strings 
($\approx 2^{N\cdot k}$) and each block-wise coded string looks almost like
if the block length were $k=1$. Therefore, the parity bit identification  
requires the knowledge of almost the full string (to within the correction 
factor of $\alpha(N,k)$ (52)).

If the block position were fixed, there would be  allowed strings
and $N$ binary tests would be sufficient to identify the parity bit.
However, the probability $p$ of success in each such test is equal to 
the sum of probabilities of occurrence of 1 or 2 or ... $k$ outcomes in the
accessible domain
\begin{equation}
p=\sum_{l=1}^{k}C_k^l \frac{1}{2^{l}}\frac{1}{2^{(k-l)}}=
1-2^{-k}.
\end{equation}
Accordingly, the probability of having $N$ outcomes (the probability of 
reliable identification of the parity bit by user B when he has only  
access to the state halves) 
\begin{equation}
P_{acc}=p^N=(1-2^{-k})^N
\end{equation}
would be high (for comparable $N$ and $k$).

The only thing we should now to do is to demonstrate that after the time 
$\tau_0+\Delta\tau \approx \tau_0$ elapses and the states become accessible 
to user B the probability of cheating by user A tends to zero. To be more 
precise, we should demonstrate that user A cannot change his mind
(modifying the chosen secret bit) after the protocol was started without being 
detected by user B with sufficiently high probability.

In the protocol, $N\cdot k$ states are sent simultaneously randomly distributed
over $N\cdot k$ channels. Possible cheating of user A is detected by user B 
with the help of a measurement described by the identity resolution of the form
\begin{equation} 
I^{\otimes^{N\cdot k} }\otimes I_{{\bf C}^2}^{\otimes^{N\cdot k } }=
\left( {\cal P}_0(f)+{\cal P}_1(f)+{\cal P}_{\bot} \right)^{\otimes^{N\cdot k}},
\end{equation} 
where
\begin{equation} 
{\cal P}_{0,1}(f)=\left( 
\frac{1}{\sqrt{2}}\int_{-\infty}^{\infty}[f(\tau)+f(\tau-\tau_0)]|\tau\rangle 
d\tau
\right)
\left( 
\frac{1}{\sqrt{2}}\int_{-\infty}^{\infty}[f^*(\tau')+f^*(\tau'-\tau_0)]
\langle\tau'| d\tau'
\right)
\otimes |e_{0,1}\rangle\langle e_{0,1}|,
\end{equation} 
\begin{displaymath}
{\cal P}_{\bot}=I \otimes I_{{\bf C}^2}-{\cal P}_0(f)-{\cal P}_1(f).
\end{displaymath}
In each of $N\cdot k$ quantum communication channels only three measurement 
outcomes are possible corresponding to 
${\cal P}_{0}(f)$, ${\cal P}_{1}(f)$, and ${\cal P}_{\bot}(f)$. If user A sends
the correct states, the different outcomes probabilities are
\begin{equation}
\mbox{Pr}\{\rho_0,0\}=\mbox{Tr}\{\rho_0{\cal P}_0(f)\}\equiv 1,
\quad
\mbox{Pr}\{\rho_1,1\}=\mbox{Tr}\{\rho_1{\cal P}_1(f)\}\equiv 1,
\end{equation}
\begin{displaymath}
\mbox{Pr}\{\rho_0,1\}=\mbox{Tr}\{\rho_0{\cal P}_1(f)\}\equiv 0,
\quad
\mbox{Pr}\{\rho_1,0\}=\mbox{Tr}\{\rho_1{\cal P}_0(f)\}\equiv 0,
\end{displaymath}
\begin{displaymath}
\mbox{Pr}\{\rho_{0,1},\bot\}=\mbox{Tr}\{\rho_{0,1}{\cal P}_{\bot}(f)\}\equiv 0,
\end{displaymath}
which means that all the outcomes should only occur with the unit probability 
in the channels ${\cal P}_{0}(f)$ and ${\cal P}_{1}(f)$ if user A employs 
the correct states. 

Any delay for a time longer than $2\Delta\tau$ introduced by user A means that 
he should employ the states which do not cover the front half of the correct
extended state, i.e. user A begins the state preparation procedure after a 
time interval exceeding $2\Delta\tau$ has already elapsed after the protocol 
was initiated. For all such states $\rho$ whose support does not cover the 
front half of the correct state the probability of the outcome in channels
${\cal P}_{0}(f)$ and ${\cal P}_{1}(f)$ does not exceed 1/2. Indeed,
\begin{equation}
\mbox{Tr}\{ \rho {\cal P}_{0,1}(f) \}=
\end{equation}
\begin{displaymath}
\frac{1}{2}\int_{-\Delta\tau}^{\Delta\tau} \int_{-\Delta\tau}^{\Delta\tau}
f(\tau) \rho(\tau,\tau') f^*(\tau') d\tau d\tau'+
\frac{1}{2}\int_{\tau_0-\Delta\tau}^{\tau_0+\Delta\tau} 
\int_{\tau_0-\Delta\tau}^{\tau_0+\Delta\tau} 
f(\tau) \rho(\tau,\tau') f^*(\tau') d\tau d\tau'\le
\end{displaymath}
\begin{displaymath}
\frac{1}{2}\cdot \frac{1}{(2\Delta\tau)^2} 
\int_{-\Delta\tau}^{\Delta\tau} \int_{-\Delta\tau}^{\Delta\tau}
|\rho(\tau,\tau')| d\tau d\tau'\le \frac{1}{2},
\quad
|\rho(\tau,\tau')|\le 1,
\end{displaymath}
if the support of $\rho$ 
\begin{displaymath}
\rho=\int_{-\infty}^{\infty} \int_{-\infty}^{\infty}
\rho(\tau,\tau') |\tau\rangle\langle \tau'| d\tau d\tau',
\quad
\mbox{Tr}\{\rho\}=\int_{-\infty}^{\infty} 
\int_{-\infty}^{\infty}\delta_{+}(\tau-\tau')\rho(\tau,\tau') d\tau d\tau'=
\int_{-\infty}^{\infty}\rho(\tau,\tau) d\tau=1,
\end{displaymath}
does not cover the front half of the correct state,
\begin{displaymath}
\mbox{supp}\rho(\tau,\tau') \cap \mbox{supp}f(\tau-\tau_0)=\emptyset.
\end{displaymath}
Hence, in the ideal communication channel any delay of the state by user A
results in the outcome probabilities in the channels ${\cal P}_{0}(f)$ and 
${\cal P}_{1}(f)$ not exceeding 1/2. To be more precise, in every individual
experiment, even for the states delayed for more than $2\Delta\tau$, 
the measurement outcome can only occur in the channels ${\cal P}_{0}(f)$, 
${\cal P}_{1}(f)$ and should never occur in the channel 
${\cal P}_{\bot}(f)$. The probability of such an outcome is 1/2. However, 
the probability of the event where in $k$ experiments all the measurement 
outcomes for delayed states occurred in the channels ${\cal P}_{0}(f)$ and 
${\cal P}_{1}(f)$ only, and thus reproduce the correct statistics 
characteristic of the non-delayed states is as small as $2^{-k}$. 
This circumstance will later be used in the protocol.

Let us now formulate the protocol itself.
               
\section{The Bit Commitment protocol for finite support states in the 
ideal communication channel.}
\begin{itemize}
\item{}
Before the protocol is started, the participant agree upon the states used
(the localization interval $\Delta\tau$ and the state shape $f(\tau)$), 
as well as the duration $\tau_0$ of the protocol (the time during which
user A retains the secret bit). In principle, time $\tau_0$ can be chosen
arbitrarily long, although this may present a difficult technical problem.
The users choose also $N$ and $k$.
\item{}
User A chooses a secret bit which is the parity bit of a string consisting of
$N$ representatives $b=\sum_{i=1}^{N}a[i,j]$, where $a[i,j]$ is bit 0 or 1, 
a representative of $j$-th block ($j$ is the number of the block consisting 
of $k$ bits). All the bits in each block are identical.
\item{}
At the moment when the protocol is started (this moment is also agreed upon
by the users in advance) user A begins to prepare $N\cdot k$ stretched states
consisting of two peaks (halves) which are allowed to propagate into $N\cdot k$ 
quantum communication channels as they are being formed. The states could also 
be sent through a single communication channel in series, although this would
substantially increase the time required for the protocol implementation.
Bearing in mind the above remarks, we assume that the channel length is zero
which actually means that user B controls only his laboratory (vicinity of
the point $x_B$) and has no control over the rest space and the communication 
channel so that user A can in principle be located just at the threshold 
of the user B' laboratory. Simultaneously, user A can control only the
vicinity of point $x_A$ where the states are prepared.

The states from different blocks $a[i,j]$ are sent through different 
channels at random.
\item{}
User B can choose any moment of time from the interval 
$\Delta\tau<\tau<\tau_0+\Delta\tau$ to start the disclosure stage
when user A should announce through a classical communication channels
which state were sent in each quantum channel and identify the quantum
channels belonging to each block.
\item{} 
User B performs the measurements described by the identity resolution
(65--67). Although the states are orthogonal (and, consequently, reliably
distinguishable), the non-local nature of the projection operators
${\cal P}_{0,1}$ implies that the reliable distinguishability (67) can only be 
achieved with the correct states if one has access to the entire state
which requires time $2\Delta\tau+\tau_0$. 
Then user B compares the results
of his measurements in each quantum channel with the data supplied to him by 
user A through the classical channel. For ideal quantum channels, the outcomes 
obtained in all the channels belonging to the same block should yield the 
identical results (all 0 or all 1). User B abandons the protocol as soon 
as he finds a discrepancy between the results of his measurements and the data
provided by user A for at least one of the quantum channel.

Note that if user A acts in an honest way (sends the correct stretched states 
at the beginning of the protocol, i.e. the parity of the string consisting of
$N\cdot k$ bits is indeed chosen at the very beginning of the protocol), no
redistribution of the quantum channels among the blocks can change the parity 
bit since otherwise the sets of odd and even strings would have common 
elements.
\item{}
If user A does not choose the bit value at the beginning of the protocol
(to be more precise, if he chooses the secret bit value after the time 
$\Delta\tau$ but, of course, before the disclosure stage, 
$\Delta\tau<\tau<\tau_0+\Delta\tau$) he will have to send the states
different from $|\psi_{0,1}\rangle$. However, for any states different from
the correct ones, the outcomes in each of the channels ${\cal P}_{0,1}$ 
will occur with the probabilities not exceeding 1/2. To modify the secret
bit, user A should delay his choice in at least one of the blocks as a whole, 
i.e. he should delay at least $k$ states. The probability of avoiding the 
detection of delay of $k$ states by user B is then $2^{-k}$ (see Eq.(68)).
\item{}
The probability for user B to have reliable information on the secret bit 
before he acquires access to the entire states does not exceed
$1/2+2^{-\alpha(N,k)N\cdot k}$ (see Eq.(62)).
\end{itemize}

Hence, the protocol allows to implement the original idea of bit commitment
scheme when one of the participants provides only part of information
analyzing which the second participant can only extract exponentially
small information on the secret bit value before the disclosure stage.
At the same time, user A cannot change the chosen secret bit after the 
protocol is started (to be more precise, the probability of undetected
modification of the chosen secret bit value after the protocol is started
is exponentially small). 

The outlined scheme allows to implement an honest protocol with
the probability not worse than $1-2^{-k}$ which is exponentially 
close to unit for large $k$.

\section{The Coin Tossing protocol for finite support states in the 
ideal communication channel.}

Although the coin tossing protocol can be constructed on the basis of
the bit commitment protocol, it is useful to formulate it explicitly.
\begin{itemize}
\item{}
Just as in the outlined bit commitment protocol, the participants A and B
agree upon the states used. When the protocol is started, each of them sends
to the other $N$ blocks containing $k$ states randomly distributed among 
$N\cdot k$ channels, the bits $b_A$ and $b_B$ chosen by users A and B,
respectively, being coded as the parity bits of the strings consisting of 
$N$ blocks. The users also agree in advance who is the winner if the
final parity bit $b=b_A\oplus b_B$ is 0 or 1.
\item{}
At an arbitrarily chosen moment of time 
$\tau$ ($-\Delta\tau<\tau<\tau_0+\Delta\tau)$ one of the users, e.g. user A,
announces for one half of all blocks through a classical channel which states 
were actually sent by him and identifies the blocks to which these states 
belong. After receiving these data, user B sends back to user A similar 
information for another half of his channels different from the channels 
disclosed by user A. Having obtained this information from B, user A
discloses which channels belong to his still unrevealed blocks and the states
that were actually sent through these channels. Then user B announces similar
information about the rest of his channels. Since the channel length
$\tau_{ch}<\tau_0$, the exchange through the classical channel can be
performed at the time when the users have access to only one half of each 
state.
\item{}
Just as in the above bit commitment protocol, the user cannot determine
the parity bit chosen by the other participant with the probability 
exceeding 1/2 until the states become fully accessible to him.
\item{} 
The users perform the measurements described by the identity resolution 
(65--67). Although the states are orthogonal (and, consequently, reliably
distinguishable), the non-local nature of the projection operators
${\cal P}_{0,1}$ implies that the reliable distinguishability (67) can only be 
achieved with the correct states if one has access to the entire state
which requires time $2\Delta\tau+\tau_0$. 
\item{}
After the time $\tau_0+\Delta\tau$ elapses and the states become
fully accessible to both users, each of them compares the results 
of his measurements in each channel with the classical information provided
by the other user. The protocol is abandoned if a discrepancy is found
in at least one channel.
\item{}
Just as in the previous protocol, the probability for each user
to obtain reliable information on the secret bit chosen by the other user
before the states become entirely accessible, does not exceed the probability
of simple guessing by an exponentially small value
$2^{-\alpha(N,k)Nk}$.

As a result, an honest parity bit (lot) $b=b_A\oplus b_B$ 
arises with the probability exponentially close to unity ($1-2^{-k}$). 

Obviously, even the correct states are sent by both users,
one of them can abort the protocol claiming the discrepancy between 
his measurement results and the classical information provided by the
other user if the arising parity bit does not suit him. However, this 
situation lies beyond the formulated problem and should be solved by
different means.
\end{itemize}
Note that sending information through the classical channel 
is necessary to avoid the cheating strategy consisting in
sending back the quantum states received from the other user
(``send back'' strategy).
For example, one of the users can send no his own states at all
and instead simply use a ``mirror'' to reflect back the states
arriving from the other user. In that case the user (say, user A) 
which wins if the final parity bit $b=b_A\oplus b_B$ is zero can 
always cheat the other user since in this situation $b_A\equiv b_B$ 
and, hence $b=b_A\oplus b_B=b_B\oplus b_B\equiv 0$. 

Disclosure through the classical channel  of the data 
on only one half of the quantum states sent by each user is also 
required to avoid the ``send back'' strategy. Had one user (say, user A) 
announced the information on all the quantum states, the second user (user B) 
could employ the ``send back'' strategy in quantum channels and simultaneously
send through another classical channel the data just received from user A
since $\tau_{ch}<\tau_0$. This strategy fails if only one half of all the
quantum states are disclosed at first. 

\section{The Bit Commitment protocol for unlocalized states in the 
ideal communication channel.}

So far we have considered the protocol employing the states with finite 
support (functions $f(\tau)\in {\cal D}(\tau)$). The set of such functions
forms a dense set in the space of functions describing the states of a free 
field (functions $f(\tau)\in {\cal J}(\tau)$). However, the field 
theory allows the states defined on the mass shell which are arbitrarily 
strongly localized and decaying with the rate arbitrarily close to the 
exponential law
($f(\tau)\propto\mbox{e}^{-\alpha\tau/\mbox{lnln}\ldots\mbox{ln}\tau}$). 
Hence one can always choose the states in such a way that the 
measurements performed over them in a finite domain on the light cone
$\tau$ yield the net outcome probability arbitrarily close to unit, i.e.
to make the contribution of the state tails at infinity arbitrarily small.
To be more precise, the states (functions $f(\tau)$) and the measurement
domain can be chosen in such a way that the probability of obtaining a 
result in the domain $\Delta(\tau)$ be
\begin{equation} 
\mbox{Pr}\{
\Delta(\tau);i,i
\}=
\mbox{Tr}\left\{ \left( \left( 
\int_{-\Delta\tau}^{\Delta\tau}
{\cal M}(d\tau)
\right)\otimes
{\cal P}_i
\right)
|\psi_i\rangle\langle\psi_i|
\right\}=
\int_{-\Delta\tau}^{\Delta\tau}
|f(\tau)|^2d\tau=1-\mbox{e}^{-\xi}\rightarrow 1,
\end{equation} 
\begin{displaymath}
|\psi_i\rangle=\int_{-\infty}^{\infty}f(\tau)|\tau\rangle d\tau
\otimes|e_i\rangle,\quad f(\tau)\in {\cal J}(\tau),\quad i=0,1.
\end{displaymath}
where $\xi$ can be arbitrarily large. Contribution from the state tails 
outside the domain $(-\Delta\tau,\Delta\tau)$ is
\begin{equation} 
\mbox{e}^{-\xi}=\int_{|\tau|>\Delta\tau}|f(\tau)|^2d\tau=
\mbox{Tr}\left\{
\left(\left(\int_{|\tau|>\Delta\tau}{\cal M}(d\tau)\right)\otimes
{\cal P}_i\right)|\psi_i\rangle\langle\psi_i|\right\}
\end{equation} 
To preserve the analogy with the case of finite support states
we shall write the stretched state with non-compact support in the form
\begin{equation} 
|\psi_{0,1}\rangle=\frac{1}{\sqrt{2}}
\int_{-\infty}^{\infty}
[f(\tau)+f(\tau-\tau_0)]|\tau\rangle d\tau
\otimes|e_{0,1}\rangle,\quad f(\tau)\in {\cal J}(\tau),
\end{equation} 
where function $f(\tau)$  ($f(\tau-\tau_0)$), just as for the one-humped
state (13) considered earlier, is strongly localized in the interval 
$(-\Delta\tau,\Delta\tau)$ ($(-\Delta\tau+\tau_0,\Delta\tau+\tau_0)$. 
The normalization condition yields
\begin{equation} 
\frac{1}{2}\int_{-\Delta\tau}^{\Delta\tau}|f(\tau)|^2 d\tau=
\frac{1}{2}\int_{-\Delta\tau+\tau_0}^{\Delta\tau+\tau_0}|f(\tau-\tau_0)|^2 d\tau=
\frac{1}{2}-\frac{1}{2}\mbox{e}^{-\xi}
\rightarrow \frac{1}{2},
\end{equation} 
and
\begin{equation} 
\frac{1}{2}\int_{\tau>|\Delta\tau|}|f(\tau)|^2 d\tau+
\frac{1}{2}\int_{\tau+\tau_0>|\Delta\tau|}|f(\tau-\tau_0)|^2 d\tau+
\end{equation} 
\begin{displaymath}
\frac{1}{2}\int_{-\infty}^{\infty}
[f^*(\tau) f(\tau-\tau_0)+f(\tau) f^*(\tau-\tau_0)]d\tau
=\mbox{e}^{-\xi}.
\end{displaymath}

Measurements performed over the stretched state in a finite window
$\Delta(\tau_0)=(-\Delta\tau,\tau_0+\Delta\tau)$ yields a result with
the probability
\begin{equation} 
\mbox{Pr}\{\Delta(\tau_0)\};i,i\}=\mbox{Tr}
\left\{
\left(\left( \int_{-\Delta\tau}^{\Delta\tau+\tau_0}{\cal M}(d\tau)\right)
\otimes {\cal P}_{0,1}\right) |\psi_{0,1}\rangle\langle\psi_{0,1}| \right\}=
1-{\cal O}(\mbox{e}^{-\xi}).
\end{equation} 
The latter term arises due to the overlapping of the tails belonging
to the two halves of the state centered at $\tau=0$ and $\tau=\tau_0$ 
and does not exceed ${\cal O}(\mbox{e}^{-\xi})$. 

Thus the statistics of measurements performed over the stretched states
should yield the results in the interval 
$(-\Delta\tau,\Delta\tau+\tau_0)$ with the probability
$1-{\cal O}(\mbox{e}^{-\xi})\rightarrow 1$ exponentially close to unit. 
The probability of obtaining a count beyond this interval does not exceed
${\cal O}(\mbox{e}^{-\xi})$ and can be made arbitrarily small by choosing
appropriate $f(\tau)$, $\Delta\tau$, and $\tau_0$.

Preparation of a delocalized state with $f(\tau)\in{\cal J}(\tau)$
formally requires an infinite time (if the state is generated by a 
point-like source) or access to the entire position space (if the state
is prepared at a specified moment of time by a delocalized source).
However, any realistic protocol should have a finite duration. To avoid
the formal problems of that kind, it is convenient to argue in the 
following way as it is usually done in similar situations. User A 
controls the neighbourhood of point $x_A$ and adiabatically
turns on the source (at $t\rightarrow-\infty$) which generates 
the vector $\psi_{0,1}(\tau_0)\rangle$ from the vacuum state. 
The source is described by the action of an $\hat{S}(\tau,-\infty)$-matrix
on the vacuum state (we do not consider the problem of
experimental realization of this source) and the produced state
\begin{equation} 
|\psi_{0,1}(\tau)\rangle=\hat{S}(\tau,-\infty)|0\rangle=
\int_{-\infty}^{\tau}
[f(\tau')+f(\tau'-\tau_0)]|\tau'\rangle) d\tau'
\otimes|e_{0,1}\rangle, 
\end{equation} 
is sent into the communication channel as it is being formed.

At the intuitive level this source can be thought of as an atomic
system (an atom) with a suitable spectrum excited by a classical field
with the appropriately chosen amplitude shape which is turned on
adiabatically and emits photons into the communication channel
(preparation of unusual one- and two-photon states is discussed, e.g.
in Ref. [35]).

User B performs measurements described by the identity resolution similar 
to Eq.(65):
\begin{equation}
{\cal P}_0(\Delta)+{\cal P}_1(\Delta)+{\cal P}_{\bot}(\Delta)=
I\otimes I_{{\bf C}^2},
\quad
I=\int_{-\infty}^{\infty}|\tau\rangle\langle\tau|d\tau,
\end{equation}
\begin{equation}
{\cal P}_{0,1}(\Delta)=
\end{equation}
\begin{displaymath}
\left(\frac{1}{\sqrt{2}} \int_{-\Delta\tau}^{\Delta\tau+\tau_0}
[ f(\tau)+f(\tau-\tau_0)]|\tau\rangle d\tau 
\right)
\left(\frac{1}{\sqrt{2}} \int_{-\Delta\tau}^{\Delta\tau+\tau_0}
[f(\tau')+f(\tau'-\tau_0)]\langle\tau'| d\tau' 
\right)
\otimes
|e_{0,1}\rangle\langle e_{0,1}|,
\end{displaymath}
and then
\begin{equation}
{\cal P}_{\bot}(\Delta)=I - {\cal P}_{0}(\Delta) -{\cal P}_{1}(\Delta).
\end{equation}
On the correct stretched states the measurement (76--78) yields the
result with the probabilities
\begin{equation}
\mbox{Tr}\{\rho_{0,1}{\cal P}_{0,1}(\Delta)\}=1-{\cal O}(\mbox{e}^{-\xi}),
\end{equation}
\begin{displaymath}
\mbox{Tr}\{\rho_{0,1}{\cal P}_{\bot}(\Delta)\}={\cal O}(\mbox{e}^{-\xi}).
\end{displaymath}
Similar to Eq.(68), for any states $\rho$ which are not entirely concentrated
in the sum of two intervals
$(-\Delta\tau,\Delta\tau)$ and $(-\Delta\tau+\tau_0,\Delta\tau+\tau_0)$, 
the measurement (76--78) yields
\begin{equation}
\int_{-\Delta\tau}^{\Delta\tau}\int_{-\Delta\tau}^{\Delta\tau}
\delta_{+}(\tau-\tau')
\rho(\tau,\tau')d\tau d\tau'=\frac{1}{2}-\frac{1}{2}\mbox{e}^{-\xi},
\end{equation}
\begin{displaymath}
\int_{-\Delta\tau+\tau_0}^{\Delta\tau+\tau_0}
\int_{-\Delta\tau+\tau_0}^{\Delta\tau+\tau_0}
\delta_{+}(\tau-\tau')
\rho(\tau,\tau')d\tau d\tau'=\frac{1}{2}-\frac{1}{2}\mbox{e}^{-\xi}.
\end{displaymath}
Hence, the delay of a state for more than $2\Delta\tau$ will result in the 
probability of obtaining a result on that state in the channel 
${\cal P}_{0,1}$ drops from almost 1 (79) to almost 1/2 (80),
\begin{equation}
\mbox{Tr}\{\rho {\cal P}_{0,1}\}=\frac{1}{2}-\frac{1}{2}\mbox{e}^{-\xi}.
\end{equation}
Accordingly, the probability in the channel ${\cal P}_{\bot}$ the probability 
rises from almost 0 (79) to almost 1/2 (80),
\begin{equation}
\mbox{Tr}\{\rho {\cal P}_{\bot}\}=\frac{1}{2}-\frac{1}{2}\mbox{e}^{-\xi}.
\end{equation}

Just as in the previous case, user A prepares $N\cdot k$ states and 
sends them into the communication channels. As long as only the halves
of the states are accessible ($\Delta\tau\le\tau\le\-\Delta\tau+\tau_0$), 
the probability for user B to obtain information on the secret parity bit 
chosen by user A does not exceed
\begin{equation}
P_c(parity)\approx \frac{1}{2}+
\left( \frac{1}{2}-{\cal O}(\mbox{e}^{-\xi}) \right)^{\alpha(N,k)Nk}.
\end{equation}
The probability of delaying the choice by user A for at least one of the blocks
consisting of $k$ bits without being detected does not exceed
\begin{equation}
\left( \frac{1}{2}-{\cal O}(\mbox{e}^{-\xi}) \right)^{k}.
\end{equation}
The probability of successfully completing the protocol (when all
$N\cdot k$ states produce the results in channels ${\cal P}_{0,1}$) is
\begin{equation}
\left( 1-{\cal O}(\mbox{e}^{-\xi}) \right)^{\alpha(N,k)Nk},
\end{equation}
can be made arbitrarily close to 1 by the appropriate choice 
of $N$, $k$, and $\xi$.

\section{Conclusions.}
Thus, the existence of maximum speed of quantum state propagation allows
to develop the relativistic quantum bit and coin tossing protocols 
explicitly implementing the original idea of the protocol where one of the
participants provides only part of information (part of quantum state) 
on the secret bit. However, the statistical nature of the measurement
procedure in quantum mechanics does not allow (at least for the proposed 
protocol) to realize the honest protocol with the unit probability. 
Nevertheless, the honest protocol can be realized with the probability
arbitrarily close to 1. In addition, the fundamental non-localizability
of the quantum field states also imposes restrictions on the probability
of the realization of the honest protocol in a finite time interval.
Nevertheless, the possibility of construction of arbitrarily strongly 
localized states allows to develop an honest protocol with the success 
probability arbitrarily close to 1 for any time $\tau_0$ (time during which
the bit secrecy is preserved).

In contrast to the non-relativistic protocols where only the structure of 
the states in the Hilbert space matters, the proposed relativistic protocols
explicitly involve the stages of the state preparation and propagation
in the space-time between the two distant users. Since the spin and helicity 
states do not exist separately from the spatial degrees of freedom of a quantum 
system, accounting for the spatial degrees of freedom extends the possibilities
for construction of quantum cryptographic protocols.

It should be emphasized once again that the protocol is based on the orthogonal 
states. A non-zero error probability in distinguishing between the two 
orthogonal states arises due to the fact that a measurement can give no outcome
at all (a photodetector will not fire, or the arrow of a classical device
will not move) if the spatial domain accessible for the measurement does not
``cover'' the entire state. Therefore, the measurement can have three outcomes:
the classical measuring device pointed to one of the channels ${\cal P}_0$ or 
${\cal P}_1$, or did not change its state at all. If the device showed any
particular result, the measured state is reliably identified. If the state
is not entirely accessible, there is a non-zero probability for the device
of not changing its state at all (showing no result at all), the larger the
inaccessible part of the measured quantum state the larger this probability.
In that case the observer can only guess what state he was dealing with
(the state identification error in this situation being 1/2).

Since the spin and helicity do not exist separately from the spatial degrees 
of freedom, the restriction of access to the position space automatically
restricts the access to the Hilbert state space. It is even possible to have 
the situation when the system state is completely inaccessible (the state
amplitude is identical zero in the domain accessible for the measurement).

It should be noted that the considered situation is different from that
discussed in Ref. [36] in connection with the analysis of Ref. [13] where 
a quantum cryptosystem based on orthogonal states was proposed. For a pair 
of orthogonal states of a composite system consisting of two subsystems
$a$ and $b$ with the state space ${\cal H}_a\otimes{\cal H}_b$
\begin{displaymath}
|\psi_0\rangle=
\alpha_0|\phi_0(a)\rangle\otimes|\phi_0(b)\rangle+
\beta_0 |\phi_1(a)\rangle\otimes|\phi_1(b)\rangle,
\end{displaymath}
\begin{displaymath}
|\psi_1\rangle=
\alpha_1 |\phi_0(a)\rangle\otimes|\phi_0(b)\rangle+
\beta_1 |\phi_1(a)\rangle\otimes|\phi_1(b)\rangle,
\end{displaymath}
where the states $\alpha_{0,1}$ and $\beta_{0,1}$ are such that the states 
$|\psi_0\rangle$ and $|\psi_1\rangle$ are orthogonal
\begin{displaymath}
\langle\psi_0|\psi_1\rangle=0.
\end{displaymath}
If only one subsystem, e.g. ${\cal H}_a$, is accessible, the states of the
other subsystem ($b$) are non-orthogonal
\begin{displaymath}
\rho_1=\mbox{Tr}_{{\cal H}_a} \{|\psi_1\rangle\langle\psi_1|\},
\quad
\rho_0=\mbox{Tr}_{{\cal H}_a} \{|\psi_0\rangle\langle\psi_0|\},
\quad
\mbox{Tr}_{{\cal H}_b}\{\rho_0\cdot\rho_1\}\neq 0,
\end{displaymath}
and, therefore, cannot be distinguished reliably. In our case the states
remain orthogonal even after the restriction to a subspace and the absence 
of reliable distinguishability arises only due to the spatio-temporal 
structure of the states.

The protocol can also be extended to a noisy channel [38] since the initial
orthogonality of the states employed allows to employ the classical codes [37].

In the proposed scheme the protocol duration time $\approx\tau_0$ 
is defined  by the effective ``extent'' of the states which for photons
can be estimated from the frequency spectrum width. The minimum attainable
spectrum width in the visible range in the ring optical fiber resonators [39]
is $\Delta\omega\approx 10$ kHz, the effective state length being 
$L \approx c/\Delta\omega=3\cdot 10^{10}/10^{4}=3\cdot 10^6$ cm (30 km). 
Accordingly, the time $\tau_0\approx 1/\Delta\omega\approx10^{-3}$ s.
Although there exist no fundamental restrictions on making time
$\tau_0$ arbitrarily long (and, respectively, $\Delta\omega$ arbitrarily 
small) this is very difficult technical problem. However, this circumstance
does not matter for the Coin Tossing protocol, since the time $\tau_0$ for 
getting an honest lot can be arbitrary. On the contrary, 
for the Bit Commitment protocol the time $\tau_0$ is an important parameter 
since it determines the time interval during which the secrecy of the chosen
bit preserved. This is a rather general situation in the experimental 
realization of various systems for transfer and processing of ``quantum 
information'' where experimental realization of the possibilities formally 
allowed by the laws of quantum mechanics today requires solution of extremely 
difficult technical problems.

It should also be noted that the large-scale information transfer systems 
are based on optical fiber where the speed of signal propagation is somewhat
lower than the speed of light in vacuum. However, this does not impose any 
restrictions since it is only necessary that the separation between 
the ``halves'' of the states used be larger than the channel length divided
by the speed of light in the optical fiber.

This work was supported by the Russian Fund for Basic Research 
(grant N 99-02-18127),  the project  ``Physical foundations of quantum
computer'' and the program ``Advanced technologies and devices of micro- 
and nanoelectronics'' (project N 02.04.5.2.40.T.50).

\end{document}